\documentclass[twocolumn,showpacs,preprintnumbers,amsmath,amssymb]{revtex4}

\usepackage{graphicx}
\usepackage{dcolumn}
\usepackage{bm}

\usepackage{graphicx}
\usepackage{amssymb}
\usepackage{epstopdf}
\newcommand{\ignore}[1]{}
\newcommand{\be}{\begin{equation}} \newcommand{\ee}{\end{equation}}
\newcommand{\ba}{\begin{eqnarray}} \newcommand{\ea}{\end{eqnarray}}
\newcommand{\nn}{\nonumber} \renewcommand{\bf}{\textbf}
\newcommand{\ra}{\rightarrow}

\renewcommand{\a}{\alpha}

\newcommand{\p}{\partial}

\def\slasha#1{\setbox0=\hbox{$#1$}#1\hskip-\wd0\hbox to\wd0{\hss\sl/\/\hss}}
\def\slashb#1{\setbox0=\hbox{$#1$}#1\hskip-\wd0\dimen0=5pt\advance
  \dimen0 by-\ht0\advance\dimen0 by\dp0\lower0.5\dimen0\hbox
  to\wd0{\hss\sl/\/\hss}}

\input{epsf}
\begin{document}

\title{ Limits on Threshold and \\ ``Sommerfeld'' Enhancements \\ in Dark Matter Annihilation}

\author{ Mihailo Backovi\'{c} and John P. Ralston}
  \affiliation{Department of Physics \& Astronomy, \\ The University of Kansas,
  Lawrence, KS 66045}

\begin{abstract} We find model-independent upper limits on rates of dark matter
 annihilation in galactic halos. The Born approximation generally fails, while exotic threshold enhancements akin to ``Sommerfeld factors''
also turn out to be baseless. The most efficient annihilation mechanism involves perturbatively small decay widths that have largely been
ignored. Widths that are very small compared to TeV mass scales suffice to cause large enhancements in the velocity averaged cross sections. Bound
state formation in weakly coupled theories produces small effects due to wave function normalizations.  Unitarity shows the Sommerfeld factor
cannot produce large enhancements of cross sections, and serves to identify where those approximations break down.

\end{abstract}




\maketitle

\section{Threshold Enhancements}

There is great interest in recent data from the PAMELA  \cite{Mocchiutti:2009sj}, FERMI \cite{Baltz:2008wd} and PPB-BETS \cite{Torii:2008xu}
experiments.  The observations suggest a significant signal in excess positron production in galactic halos, as long suggested by the HEAT
\cite{Barwick} and ATIC \cite{chang} experiments.  Possible explanations range from exotic mechanisms \cite{exotic}, uncertain features of pulsars
\cite{pulsars}, to dark matter decays \cite{decays} and dark matter annihilation \cite{Barger:2009yt}.

In considering annihilation there are puzzles from comparing predictions of relic densities with rates of particle production in the current era.
This has led to invoking more or less exotic threshold enhancements under the catch-phrase of ``Sommerfeld factors'' \cite{Hisano:2003ec,arkani}.

One reason to appeal to a Sommerfeld factor is to boost cross sections of TeV-scale particles well above Born-level estimates. We find the
starting point of Born-level cross sections is not a good approximation for much different reasons.  Basic facts of finite width particle physics
substantially revise estimates of annihilation rates in galactic halos. We find that annihilation of TeV-scale dark matter with typical
electroweak couplings can actually saturate unitarity limits over the observable range. We also obtain upper limits to halo annihilation rates
that do not depend on fine details of the dark matter velocity distribution.

Non-relativistic scattering amplitudes can be classified by their analytic properties in the complex momentum plane. Stable bound states are
described by poles on the positive imaginary axis. It follows that stable bound states produce no remarkable enhancement of annihilation rates in
the physical region of real momentum $k$. {\it Metastable particles} or {\it resonances}, described by poles of finite width, are in no way
comparable with stable bound states, because everything observable (and potentially large) is a strong function of the width.

Unless one is considering an absolutely stable intermediate state, {\it all intermediate states in particle physics have a finite lifetime.}
Pursuing the consequences of finite lifetimes with galactic halo kinematics re-directs attention from exotic mechanisms to ordinary physics. There
are two salient cases. If the width of an intermediate annihilation state is limited by the initial state velocity $v$, then the peak of the cross
section goes like $1/v^{2}$. This case produces the largest reaction rates in halos, and most conservative bounds.  If the width of the
intermediate state is constant, the peak of the cross section goes like $1/v$.  In these and intermediate cases the peak cross section actually
dominates the entire halo velocity distribution for a surprisingly broad range of dark matter parameters. As a result, our more conservative
bounds merge smoothly with reasonable estimates predicting surprisingly large rates.

\section{Breit-Wigner Formulas}

Relic particles trapped in galactic halos will be non-relativistic, with velocities $v \sim 10^{-3}$.  There are several distinctly different
non-relativistic ``Breit-Wigner '' formulas. Most Breit-Wigner cross sections $\sigma_{res}$ can be cast into the form \ba \sigma_{res} &=& { 4
\pi  v^N \over k^{2}} {(\Gamma /2)^{2} B_{i}B_{f} \over (E- E_{res} )^{2}+(\Gamma /2)^{2}} = { 4 \pi v^N \over k^{2}} \, BW(\Gamma, \, E_{res}).
\nn
\\  \label{bw} \ea Here $B_i$ and $B_f$ are the branching fractions to the initial and final state, and $k$ is the momentum of an initial state
particle in the center of mass frame. Different values of the parameter $N =0, \, 1$ distinguish two classic limits:

{\it Phase Space Limited Case, $N=0$:} It is common for $2 \ra 2$ non-relativistic physics to be quasi-elastic. In particular, the {\it final}
state phase space may be severely limited by the {\it initial} state velocity $v$. Ignoring spin and matrix elements, the Lorentz-invariant phase
space integral $LIPS$ for two particles of momentum $p_f$, $p_fÕ$ and mass $m_{f}$: is  \ba LIPS= \int { d^{3} p_f \over 2 p_f^0}  {d^{3} p_fÕ
\over 2 p_f^0Õ} \delta^4(Q-p_f-p_fÕ) \nn \\ = 2 \pi \sqrt{1-4m_f^2/Q^2} =2 \pi v_f. \label{ps} \ea Here $v_{f}$ is the final state velocity of
either particle in the CM frame. When initial and final state masses are comparable, and the 2-body states dominate, the total width $\Gamma \sim
\kappa v_{f} \sim \kappa v$, where $\kappa$ absorbs coupling constants and matrix elements. Incorporating the explicit velocity dependence with an
$s$-channel propagator leads to Eq. \ref{bw} with $N=0$.  Note that the peak of the cross section scales like $1/(m^2v^{2})$, making this case
potentially capable of saturating elastic unitarity bounds.

{\it Relativistic Phase Space Case, $N=1$:} Anihillation may also proceed to final states which are ultra-relativistic. Then the square root in
Eq. \ref{ps} approaches 1, and the partial width $\Gamma_{f} \sim constant$ in this limit. Any other kinematic situation where $Q^2/m_f^2$ goes to
a finite constant as $v \ra 0$ will produce the same outcome. This includes the  ``exoergic'' resonances long known in low-energy nuclear physics,
and associated with the "$1/v$ law" of low energy cross sections.  These cross sections do not increase as fast as unitarity would allow as $v \ra
0$.

The difference between $1/v$ and $1/v^2$ velocity dependence is dramatic.  Yet is only part of the story, because  resonances may produce large
cross sections either way.  For example, neutron absorption cross sections on Gadolinium-157 exceeding {\it one hundred million barns} have been
observed. \cite{gadolinium}. This comes in the seemingly ÒmildÓ $1/v$ case not impinging on a unitarity limit. The experimental stunt simply
exploits neutrons with grossly small velocities of order 3 meters per second. In much the same way, galactic halo velocities of order $10^{-3}$
are grossly small on the scale of particle physics.  The combination of low speed halo kinematics and very ordinary widths produces surprisingly
large enhancements.

\section{General Limits on the Velocity Averaged Breit-Wigner Cross Section}
The halo annihilation rate via a single $s-$wave resonance is governed by the velocity-weighted cross section $\langle \,  \sigma v  \,
\rangle_{res}$: \ba \langle \,  \sigma v \, \rangle_{res} &=& \int dv \, v { 4 \pi v^N \over m_{X}^{2}v^{2} } \nn \\ & \times &
{(\Gamma/2)^{2}B_{i}B_{f} \over (m_{X}v^{2}/2- m_{X} v_{res}^{2}/2 )^{2}+(\Gamma/2)^{2}}\Phi_{halo}(v) . \nn \ea Here $\Phi_{halo}(v)=dN/dv^{3}$
is the normalized dark matter relative velocity distribution, assumed from astrophysics to be a smooth function on the scale of 100-500 $km/s$. In
an isothermal halo model the velocity distribution is in equilibrium,  \ba {dN \over d^{3}k d^{3}x} &=& {constant \,\over E_{0}}e^{-E/2E_{0}}; \nn
\\ {dN \over dv } &=& 4 \pi { v^{2} \over (2 \pi v_{0}^{2})^{3/2}} e^{-v^{2}/2v_{0}^{2}}. \label{equil}\ea While the actual velocity distribution is
uncertain the phase space factors of $v^{2}$ are general. The isothermal halo will illustrate the method, but none of our upper bounds depend
on it.

The rate $\langle \, \sigma_{res}v  \,  \rangle_{res} $ is a function of $E_{0}$, $E_{res}$, $\Gamma$ and $m_{X}$. If other scales are expressed
in units of $m_{X} \sim TeV$ the conjunction of several rapidly varying functions makes analysis troublesome, as noted by Griest and Seckel
\cite{griest}. However in the present universe the halo energy $m_{X}v_{0}^{2}/2 \sim 10^{-6}m_{X}$ is rather small on particle physics scales. It
is natural to rescale variables in units of the halo characteristic energy,   defining\ba \gamma_{0} ={\Gamma \over 2 E_{0}} ;\:\:\:\:\:
\delta_{0} = {E_{res} \over E_{0}} .\nn \ea Assuming the equilibrium distribution, some algebra gives \ba \langle \,  \sigma v \,  \rangle_{res}
&=& {2^{2-N} (2 \pi)^{\frac{N+1}{2}} v^{N-1}_0\over m_{X}^{2}} I_N (\gamma_{0}, \, \delta_{0}), \label{rate} \ea where \ba I_N(\gamma_{0}, \,
\delta_{0}) = \frac{1}{2^{1-N}(2 \pi)^{N/2}}\int_{0}^{\infty} dz z^{\frac{N}{2}} \,{ \gamma_{0}^{2} e^{-z/2}\over (
z-\delta_{0})^{2}+\gamma_{0}^{2} }. \label{inty} \ea

\begin{figure}
\begin{center}
\includegraphics[width=2.5in]{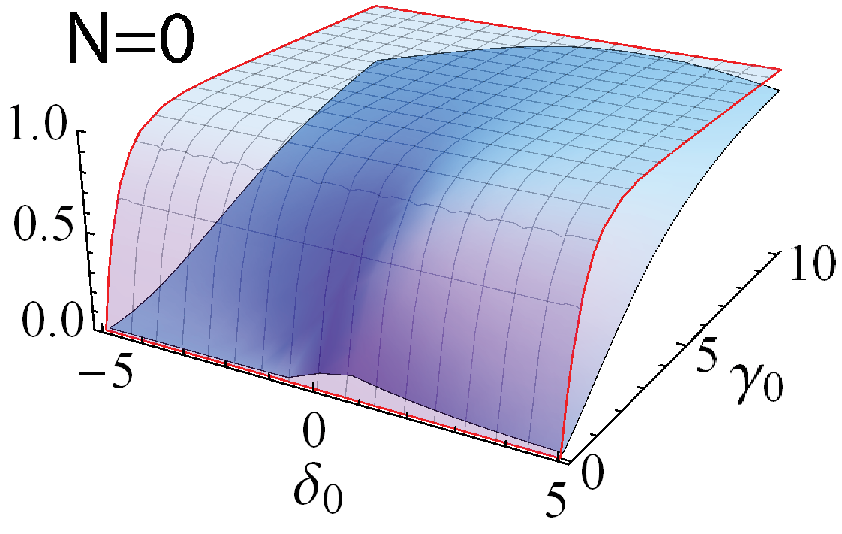}
\includegraphics[width=2.5in]{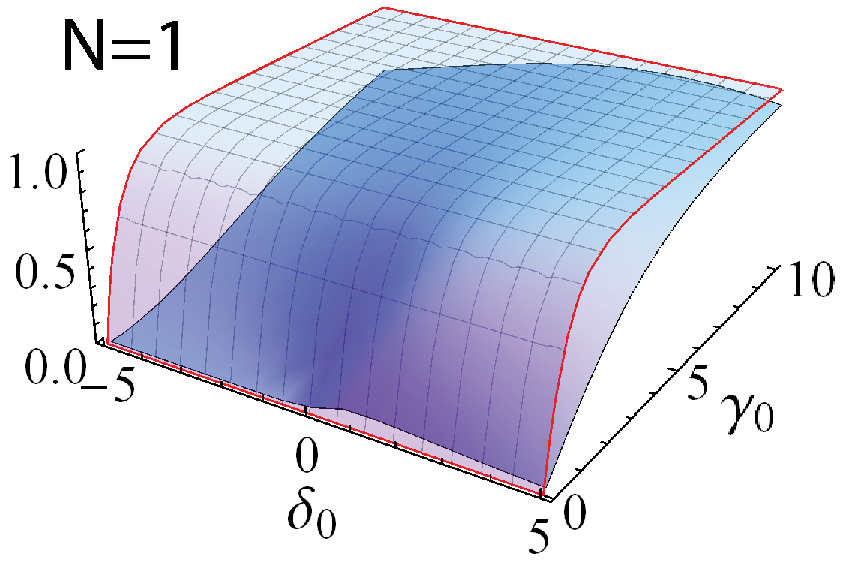}
\caption{ The integral $I_N(\gamma_{0}, \, \delta_{0})$ (dark shaded) and upper limits cited in the text (transparent mesh).  By Eq. \ref{rate}
the rate constant is related via $\langle \sigma v \rangle_{res} \sim v^{N-1}_0 I_N/ m_{X}^{2}$.  }
\label{fig:SigmaVInt.eps}
\end{center}
\end{figure}

Note that $I_N(\gamma_{0}, \, \delta_{0})$ is analytic for all $\gamma_{0}>0$ and $\delta_{0}$ regardless of the sign of $\delta_{0}$.  It can be
computed exactly in terms of Exponential Integral ($Ei$) functions. We found it more useful to observe that $I_N(\gamma_{0}, \, \delta_{0})$ has
certain {\it absolute upper limits} for all possible values of $\gamma_{0}>0$ and $\delta$.  Consider the derivative $\p I_0(\gamma_{0}, \,
\delta_{0})/\p \gamma_{0}$: \ba { \p I_0(\gamma_{0}, \, \delta_{0}) \over \p \gamma_{0} } = \int dz\, \frac{2 \gamma_{0}
(\delta_{0}-z)^2}{\left(\gamma_{0}^2+(\delta_{0}-z)^2\right)^2}e^{-z} . \label{ineq} \ea Since the integrand above is positive definite, the
integral achieves its maximum at $\gamma_{0} \ra \infty$. For $\gamma_{0}>>1$ the integration becomes trivial, yielding $I_N(\gamma_{0}, \,
\delta_{0}) \leq 1$. A stronger limit notes the integrand of Eq. \ref{inty} is cut off for $z \lesssim \gamma_{0}$ when $\gamma_{0} \lesssim 1, \,
\delta_{0}\lesssim 1$, implying $I_N(\gamma_{0}, \, \delta_{0}) \lesssim 1-e^{- C \, \gamma_{0}}$, where $C$ is a constant. Numerical work shows
that for all parameters \ba I_0(\gamma_{0}, \, \delta_{0}) \leq 1-e^{-\frac{\pi}{2} \gamma_{0}}, \\ \nn
               I_1(\gamma_{0}, \, \delta_{0}) \leq 1-e^{-\frac{\pi}{4} \gamma_{0}}. \nn \ea
These are close to equality for positive $\delta_{0} <<1$. Figure \ref{fig:SigmaVInt.eps} shows a plot of $I_N(\gamma_{0}, \, \delta_{0})$for a
wide range of $\gamma_{0}, \, \delta_{0}$ and how the integral approaches the upper bound.

The positivity property of Eq. \ref{ineq} holds for all halo distributions. The upper limit $BW \ra 1$ produces a
universal inequality: \ba \langle \, \sigma v \,  \rangle_{res} < {4 \pi \langle \,  1/v^{1-N}  \,  \rangle  \over m_{X}^{2} } . \label{inequal}\ea The
expected value $\langle \,  1/ v  \,  \rangle  $ is relative to the distribution $\Phi_{halo}(v) $, not $dN/dv$.  If the equilibrium distribution
is assumed, then \be \langle \, \sigma v \, \rangle_{res} < {2^{2-N}
(2 \pi)^{\frac{N+1}{2}} v^{N-1}_0\over m_{X}^{2}} ( 1-e^{- \pi \gamma_{0} /2^{N+1} }) \label{g0bound}\ee

The result is a possible significant enhancement factor ($EF$) (``boost factor'') for annihilation rates.   The enhancement factor is defined
relative to a typical Born approximation $\sigma_{Born} =4 \pi \a_{X}^{2}/m_{X}^{2}v^{2-N}$: \ba EF = {\langle \, \sigma v \,  \rangle_{res} \over
\langle \, \sigma v \,  \rangle_{Born}} \lesssim {1 \over \a_{X}^{2}}. \label{ef} \ea

Note that the upper limit does not depend on the position of the resonance nor on any halo properties.

\subsubsection{$N=0$ Enhancement Factors}

For $N=0$, Eq. \ref{ef} leads to substantial enhancements approaching the unitarity bound when the fundamental width $\Gamma$ is large enough.
Obtaining a ``large enough'' width from a weakly coupled theory might appear special. Yet remember that halo annihilations are driven by the width
in units of the rather small scale $E_{0} \sim 10^{-6}m_{X}$.  For TeV-scale dark matter a width $\Gamma \gtrsim $ MeV is large enough to dominate
the halo width and make $BW(\Gamma, \, E_{res}) \sim 1$. Recall that the $J/\psi$ has a width of order 0.1 MeV and is exceedingly ``narrow''. For
an elementary particle on any mass scale of GeV-TeV {\it not to have} widths exceeding $10^{-6}m_{X}$ requires special conspiracies or selection
rules.

\begin{figure}[!]
\begin{center}
\includegraphics[width=2.5in]{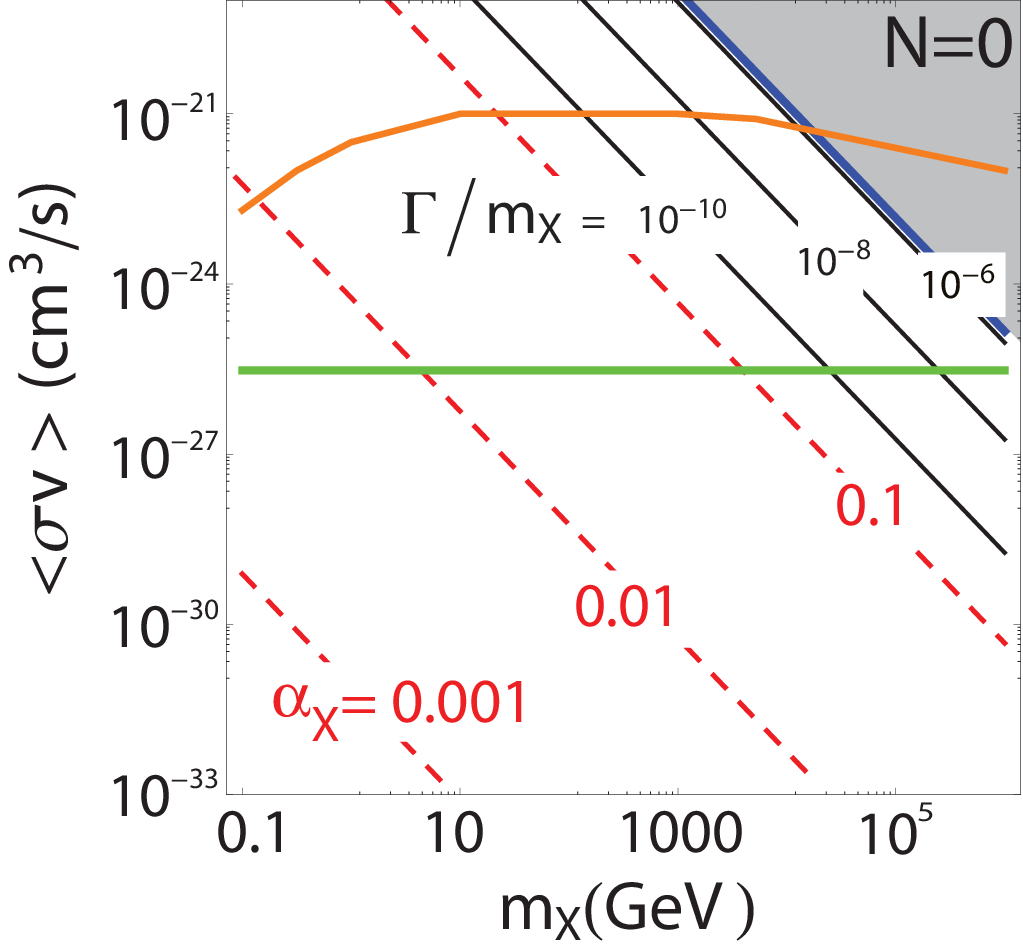}
\includegraphics[width=2.5in]{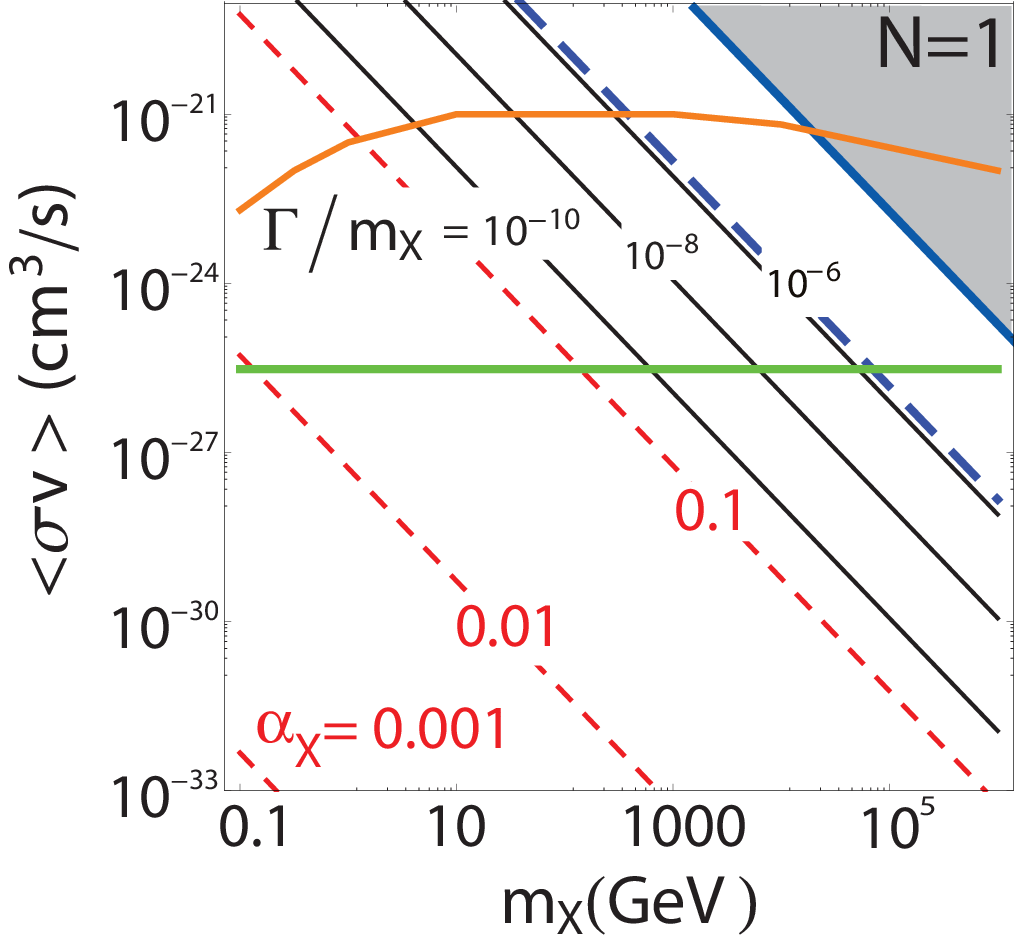}
\caption{ Upper limits (diagonal lines) of resonantly enhanced annihilation rate $\langle \,  \sigma v  \,  \rangle_{res}$ in the isothermal halo
distribution.  Solid curves (black online) are computed with fixed $ \Gamma/m_{X}$. Gray triangle in upper right is the unitarity bound. The thick
dashed curve (blue online) is the maximum value for the cross section for $N=1$. Thin dashed curves (red online) show $\langle \,  \sigma v  \,
\rangle_{res}$ computed for bound state processes using $\Gamma = \a_{X}^{5}m_{X}/2$ and $E_{res} = -m_{X}\a_{X}^{2}/4$.  Middle curve (orange
online) is the neutrino-based upper limit of Ref. \cite{Beacom:2006tt}. Horizontal line (green online) is a conventional lower bound $\langle
\sigma v \rangle \sim 3 \times 10^{-26} cm^{3}/s$. } \label{fig:ResLimits2.eps}
\end{center}
\end{figure}

Figure \ref{fig:ResLimits2.eps} shows that even a tiny value of $\Gamma/m_{X} \sim 10^{-8}$ can produce  rates much larger than the oft-cited
value $\langle \, \sigma v \,  \rangle  \sim 3 \times 10^{-26} cm^{3}/s$.  It is a new insight that
merely including physics of widths tends to saturate unitarity bounds in halo annihilation.

\subsubsection{$N=1$ Enhancement Factors}

\begin{figure}[!]
\begin{center}
\includegraphics[width=2.5in]{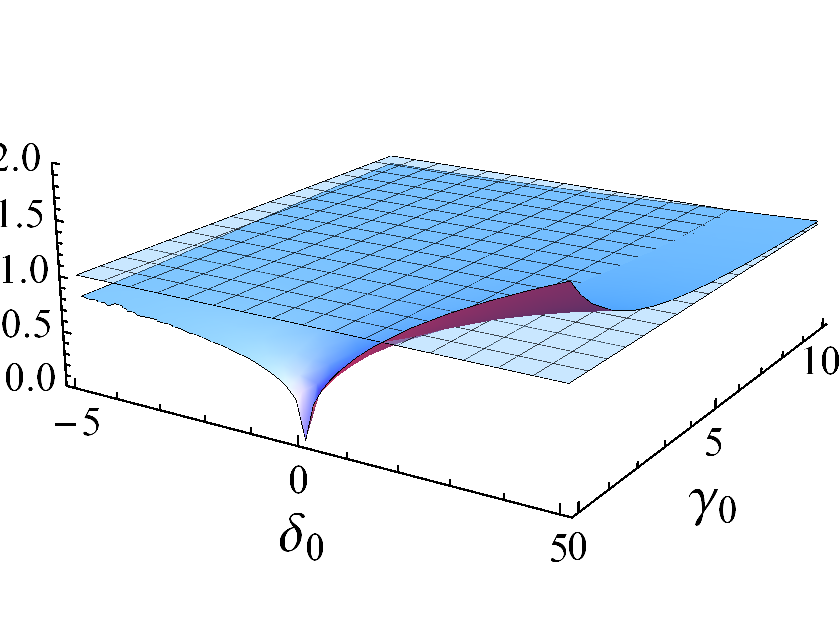}
\caption{ Ratio of $I_1/I_0$ (shaded area) compared to the uniform value of 1 (mesh)\label{integralratios}}
\end{center}
\end{figure}

Equation \ref{inequal} highlights a factor of $\langle 1/v \rangle$  absent with a relativistic phase space ($N=1$).  To a first approximation the
ratio of the $N=1$ case relative to the $N=0$ case is of $O(v_0)$. This is made more precise using Figure \ref{integralratios}, which shows a plot
of the calculated ratio of integrals $I_1/I_0$ that remains.  This ratio is of order unity for most of the parameter space, except the regions
where $\gamma_0 << 1$.  Once again, only when widths are very tiny do resonance widths {\it not} tend to swamp the halo distribution.

While representing stronger limits, the bottom panel of Figure \ref{fig:ResLimits2.eps} again shows significant enhancements over a broad range of
parameters of current interest. The difference between $I_1$ and $I_0$ tends to disappear whenever  $\Gamma/m_X$ is not exceptionally small. In
the next Section we turn to the metastable bound state case, which does happen to exhibit exceptionally small widths on general grounds.

\subsection {\it Metastable Bound States, and Narrow Resonances}

The case of annihilation passing through intermediate metastable bound states has generated great interest. This case is different and deserves a
separate discussion. Suppose dark matter interacts with a light messenger particle of mass $\mu$, with coupling-squared $\a_{X}$. If the
interaction is attractive, which is readily arranged for particular spins, then non-relativistic physics predicts {\it there is always a bound
state} for sufficiently large coupling. The conditions are \ba \a_{X} \gtrsim \kappa { \mu \over m_{X}}, \nn \ea where $\kappa$ is a constant of
order one. The demonstration is an easy variational calculation using a ground state Hydrogenic wave function. A helpful discussion is also given
in Ref.  \cite{Shepherd:2009sa}. For parameters $m_{X} \sim TeV$ and $\mu \sim GeV$ bound state formation needs $\a_{X} \gtrsim 10^{-3}$, which is
well within the electroweak-scale couplings of most models.

Yet just as above, everything about any significant enhancement depends strongly on the {\it width}, and won't proceed without it. To estimate
widths, first note that bound states are spatially large for small coupling constant $\a_{X}$. The size of a weakly coupled bound state is roughly
estimated by the ``Bohr radius'' $a_{0}$, where \ba a_0 \sim 1/m_{X}\a_{X}. \nn \ea Similarly, the binding energy is $E_{res} \sim
m_{X}\a_{X}^{2}$. Next recall that  the Schroedinger wave function at the origin $\psi(0)$ determines the width via $\Gamma \sim |\psi(0)|^{2}
\sigma_c  $ where $\sigma_c$ is a continuum cross section.

The wave function at the origin is set by the inverse of the size of the bound state: \ba |\psi(0)|^{2} \sim a_{0}^{-3}\sim \a_{X}^{3}. \nn \ea
The continuum annihilation cross section $\sigma_{c} \sim \alpha_{X}^{1+A}$  for $A>0$ depending on the model.  For reference the annihilation
rates of ortho (para) positronium via three (two) photons go like $\a_{em}^{6}$($\a_{em}^{5}$). Thus bound state widths follow a general pattern
\ba \Gamma \sim \a_{X}^{4+A}m_{X} \lesssim 10^{-8} m_{X}. \nn \ea The right hand side is a fair upper limit for $\a_{X}\sim 10^{-2}$. Restricted
phase space factors and branching ratios can only reduce this. Comparing $E_{0}\sim 10^{-6}m_{X}$, we find that $\gamma_{0}<<1$ is by far the
generic case for annihilation from a bound state.  As a consistency check, consider the definite case of spin-1/2 dark matter interacting with
vector particles. Nature has already done this calculation with the $J/\psi$ decay via gluons, which has $\Gamma_{J/\psi}/m_{J/\psi} \lesssim
10^{-4}$. The $J/\psi$ is sufficiently heavy that the perturbative phase space factors are driven by dimensional analysis, as expected for
TeV-scale physics. The raw $J/\psi$ ratio needs to be re-scaled by $(\a_{X}/\a_{s} )^{4+A} \sim 10^{-4}$, which gives satisfactory agreement.

When $\gamma_{0}<<1$ it is a good approximation to replace $BW(\Gamma, \, E_{res}) \sim \pi (\Gamma/{2}) \delta(E-E_{res})$. A short calculation
then gives \ba EF(\gamma_{0}<<1) = { \pi \Gamma/2 \over \a_{X}^{2} m_{X} v^{2-N}_{res}} { \Phi_{halo}(v_{res}) \over \langle \, {1 \over v^{1-N}} \,
\rangle} , \label{gamlim}\ea where $E_{res} =m_{X}v_{res}^{2}/2$. This formula has no singularity as $v_{res} \ra 0$ because $\Phi_{halo}(v_{res})
\sim v_{res}^{2}$ has compensating factors from phase space (Eq. \ref{equil}). If a metastable bound state resonance lies {\it above threshold} in
an expected electroweak range the effects are quite small. Taking $E_{res} =m_{X}\a_{X}^{2}/2 \sim 10^{-4}m_{X}$, and the equilibrium halo model
with scale $v_{0}=10^{-3}$, the factor $e^{-E_{res}/E_{0}} \sim e^{-100}$ is too small to consider further. When the resonance is {\it below
threshold} it must have width $\Gamma \gtrsim |E_{res}|$ to intrude into the physical region. Since $\Gamma$ is proportional to several powers of
$\a_{X}$ compared to $E_{res}$ this case can also be set aside. If $E_{res} \ra 0$ with $\Gamma >>E_{res}$ is contemplated, it implies the decay
time scale is much less than a binding (orbital) time scale, which is not consistent with bound states forming in the first place.

An exponentially small suppression can be avoided by adjusting the binding into the range probed by the halo velocity. For example choose $\a_{X}
\sim 10^{-3}$. This device rapidly loses consistency because the bound state criterion $\a_{X} \gtrsim \mu/m_{X}$ needs couplings not too small.
If a bound state is tuned to the vicinity of the peak, then the halo factors will be order unity. Meanwhile there remains in Eq. \ref{gamlim} an
overall factor of $\Gamma/(m_{X} \a_{X}^{2})<<1$.  Figure \ref{fig:ResLimits2.eps} compares the upper limits from annihilation of continuum
processes ($\gamma_{0} \gtrsim 1$ generically) to processes proceeding via the bound state ($\gamma_{0} \lesssim  10^{6}\a_{X}^{5}$) using the
isothermal halo and conservative values $B_{i}B_{f} \ra 1$. Viable enhancement mechanisms should also respect the neutrino-based bounds of Mack,
Beacom and Bell \cite{Beacom:2006tt} included in the Figure. In case of $N=0$, a bound state could cause large cross section enhancements, but
only for couplings $\alpha_X \geq 0.1$ which are beyond the stable perturbative regime. In case of $N=1$ the limits for bound states are even
tighter.

Figure \ref{fig:ResLimits2.eps} shows that a single bound state with perturbative couplings has no chance of causing significant enhancements.
Except for strong coupling, there is {\it no dynamical mechanism to generate large enhancement factors from non-relativistic bound
state resonances in the current universe.} The conclusion does not depend on the spin or quantum numbers of new physics, and is too strong to
escape by adding up several resonances, unless they are so numerous their numbers alone overcome small couplings, as for $KK$ modes.

\subsubsection{Breit Wigner Effects on Relic Abundance}

Relic abundance is a different topic than halo annihilation.  Ibe, Murayama, and Yanagida \cite{Ibe:2008ye},and Guo and Wu \cite{Guo:2009aj} have
calculated thermal evolution for the case of a narrow state close to threshold.  Their model cross section is essentially equivalent to our $N=1$.
The resonance position is close enough to threshold for its effects to overlap into the physical region during relic evolution. They find that
even a tiny ratio of width to resonance invariant mass, denoted by $\gamma=\Gamma/M_{res} \lesssim 10^{-3}$, produces significant effects on relic
densities compared to traditional constant cross section approximations.  The sense of this effect causes a relative decrease in annihilation
rates in the early universe, which tends to leave too much relic.  To keep the relic density $\Omega_{X}$ fixed, avoiding over-closure, Refs.
\cite{Ibe:2008ye,Guo:2009aj} introduce ``boost factors'' to correct the normalization parameters of the cross section $\sigma_{0} \sim
B_{i}B_{f}/m_{X}^{2}$.  When those boost factors are applied directly to halo annihilation, they develop much larger cross sections than the
well-known cosmological value $\sigma_{0}\sim 10^{-9} \, GeV^{-2}$, which might be relevant to the PAMELA-ATIC observations.

However, in general it is not possible to go from the relic calculation to the halo calculation directly in this manner. The halo annihilation
rate $<\sigma v>$ has a new and separate sensitivity that is {\it a priori} disconnected from relic calculations.  The halo estimates are driven
by the new parameter $\gamma_{0} = \gamma \,M_{res}/2E_{0}$. As long as $\gamma_0>> 1$, the upper bounds on the halo annihilation cross section
will be saturated. This key feature is conceptually absent if the halo annihilation cross section is simply re-scaled by factors invoked for relic
evolution.   Thus the reported ``boost factors'' of the relic calculations do not take into account the Breit-Wigner effects on halo annihilation
we have found.  This explains why the suggestion \cite{Ibe:2008ye,Guo:2009aj} that very small $\Gamma/M_{res}$ is necessary or tends to enhance
halo rates is not general, and appears different from our conclusion. It is clearly possible to find models and parameter regions where both, or
neither of the correct relic density and halo enhancement phenomena can be accommodated.

We note that relic densities are also subject to many uncertainties of galaxy formation and the other boost factors representing ``clumpiness''.
For purposes of confronting experimental data, it seems best to separate the problems of halo annihilation and relic evolution entirely, despite
mathematical similarities in how they are calculated.

\section{Sommerfeld Factors}

We have shown that Breit-Wigner width effects of typical particle physics type can be surprisingly large, while bound state effects have little
chance to compete. {\it Sommerfeld factors} have also been claimed as a mechanism to produce large enhancements not involving particle widths
\cite{arkani}.

Given an $s-$wave cross section $\sigma^{0}$, which has been computed in the plane wave basis, the Sommerfeld-based recipe to include Coulomb wave
effects is to make a replacement \ba \sigma^{0} \ra \sigma^{0} S(v, \, \a); \nn \\ S(v, \, \a) = {\a \over v }{2 \pi \over 1- e^{-2 \pi \a /v} }.
\nn \ea Here $\a$ is the fine structure constant. Since cross sections contain many other terms of different orders in $\a^{j}/v^{k}$, together
with logarithmic type dependence, the recipe is an approximation by {\it re-summation} of selected contributions  \cite{milton}.

\subsubsection{Motivation for Re-summation}

Non-relativistic $QED$ has complicated logarithmic and power-behaved infrared singularities. Singular terms must be summed or controlled in some
way to avoid upsetting perturbation theory. There are reasons to believe that the leading singularities appear \footnote{Many papers, e.g. Ref.
 \cite{brodsky} quote results without derivation or approximations. We have not actually found a proof in the literature} order by order as a
series in $\a^{j}/v^{j}$.  Evidently such a series is summed in $S(v, \, \a) = S(\a/v)$.

Sub-leading terms are dropped in any re-summation - for example, a term of order $\a^{j}/v^{k}$ with $j>k$ is sub-leading as $v\ra 0$. The fact
that infinitely many sub-leading terms exist comes from the fact that it is always possible to add a photon exchange loop to any diagram, and all
loops have some integration region {\it not} singular as $ v \ra 0$.

The purpose of re-summation is to extend the reach of perturbation theory into the difficult, non-perturbative regime.  Some examples illustrate
typical limitations. Expand $S(v, \, \a)\sim 1 +\pi(\a/v)+\pi^{2}\a^{2}/ 3 v^{2}+...$.  For $\a=10^{-2}$, and $v=5 \times 10^{-2}$ the third term
is $\pi^{2}/75   \sim 0.13$. It happens to be larger than a typical non-singular term of order $\a$; retaining it is well-motivated for these
kinematics. The series is also stable in this regime: a high order term $(\pi \a/v)^{10} \sim 0.0096$ is small.  Yet at the smaller velocity
$v=5/1000$, the sub-leading correction $\a^{11}(\pi /v)^{10}  \ra 958956.0$ is not included, is hardly small, and the leading-order re-summation
fails. It is the nature of such re-summation that self-consistency breaks down in the region $\a /v \gtrsim 1$, exactly where $S(v, \, \a) \gtrsim
1$.

If one believes a re-summation recipe might generate corrections of order 30-50\%, say, there's seldom any reason to invoke it for factors of
``10'' or more.  Yet recent treatments of dark matter annihilation have imagined the Sommerfeld factor to be very general.  It has been held
responsible for extremely large enhancement factors of $S>>10$, while also coming from any generic interaction involving light Yukawa particles
\cite{recents}. The perception comes from a practice of citing continuum Coulomb normalization factors $|\psi_{C}|^{2}(0) $.  Since Coulomb
normalizations are known exactly, the procedure has been thought to be ``exact.''

We have traced early literature to find several logical and historical contradictions.  Guth and Mullin highlighted the approximations in 1951
\cite{guthMullin}.  {\it In lowest order approximation}, but while using Coulomb wave functions $\psi_{C \, i}$ for a basis, one encounters matrix
elements $M$ of the form \ba M =\int d^{3} x \, \psi_{C \, 2}^{*} V \psi_{C \, 1} \nn.  \ea  Insert complete sets of momentum eigenstates $|k
\rangle$: \ba M=\int d^{3}k\, d^{3}k'  \, \psi_{C \, 2 }^{*}(k)V_{kk'}\psi_{C \, 1}(k'). \nn \ea  The Coulomb wave functions are sharply peaked in
the vicinity of certain momenta $k_{1}$, $k_{2}$, identified by taking the limit $\a =0$.  Assume the plane-wave matrix elements $V_{k_{1}, \,
k_{2}} = \langle k_{1}|V|k_{2} \rangle$ are relatively smooth functions of momentum transfer. Make a rough approximation moving $V_{k_{1}k_{2}'}$
outside the integral:  \ba M & \ra &  V_{k_{1}, \, k_{2}} \int d^{3}k \, \psi_{C \, 2}^{*}(k) \int d^{3}k' \, \psi_{C \, 1}(k), \nn \\  & \ra & \
V_{k_{1}, \, k_{2}} \psi_{C \, 2}^{*}(x=0) \psi_{C \, 1} (x=0). \label{resum} \ea In the last line $\psi_{C } (x=0)$ appears as the
coordinate-space wave function at the origin, ``improving'' the plane wave calculation. Inserting the analytically known normalization of
$|\psi_{C } |^{2}$ then produces the factor $S(v, \, \a)$ for the cross section.

The operations of separating the collision and wave function integration into products is one of {\it leading power factorization}. It is used in
$QCD$ calculations separating ``hard'' and ``soft'' regions of perturbation theory, but in that case while attempting to be systematic. Careful
work with positronium annihilation \cite{harris} does not use the factorized approximation.  Instead, reference to re-summation is made after the
full calculations are carried out.  An early work \cite{harris}  on one-loop corrections to positronium decay states that``Coulomb effects are
included by this (factored) method to all orders in $e^{2}$, {\it though only, of course, approximately}.'' (Italics are ours.)

What did Sommerfeld actually do?  We consulted his 1931 article, in German, to see it introduced exact Coulomb wave functions to calculate
bremsstrahlung, while it never suggested factorization.  It is a {\it tour de force} of early quantum theory; consulting it for a renormalization
factor actually perpetuates a normalization mistake.  Cross sections are defined by {\it ratios relative to a flux computed with a given
normalization}. The overall normalization of physical states cancels out in total cross sections: and so Eq. \ref{resum} is not only approximate,
it is incomplete.

Elwert's 1939 dissertation \cite{elwerthaug} recognized this, as as did Guth in 1941 \cite{guth}. These papers abandoned Sommerfeld's calculation
and used the {\it ratio} of two in- and out- Coulomb factors as an approximate factorized ansatz. ``Elwert factors'' are used in atomic and
molecular physics to cancel spurious pre-factors going like $v$ from other approximations, but only when their effects are not too large.
Experimental confrontation of the Elwert factor finds errors of relative order unity in the region the factors are of order unity
\cite{pratt,elwerthaug}.  Elwert and collaborators find this kind of breakdown reasonable \cite{elwerthaug}. In no event are very large
corrections ever credited.

\subsubsection{ Multiplicative Factors Must Fail}

In retrospect, we find the concept behind generating singularities via multiplicative factors questionable on general grounds.

A general scattering amplitude has the partial wave expansion \ba f(\theta, \, k) &=& {1\over k} \sum_{l} \, (2l+1)f_{l}(k)P_{l}(cos \theta).\nn
\ea  For each partial wave cross section $\sigma_{l}$ of angular momentum $l$, elastic unitary gives the upper limit \ba \sigma_{l} = {4 \pi
(2l+1) |f_{l}(k)|^{2}\over k^{2}} \leq {4 \pi (2l+1) \over k^{2}} . \nn \ea This summarizes the unitarity bound of Ref.  \cite{Hui}. Since each
partial wave has a finite cross section, {\it no partial wave can possibly have a singularity}. ``Improving'' the $s$-wave cross section - or any
particular partial wave cross section - by singular terms of order  $\a_{X}/v$ then contradicts unitarity for $\a_{X}/v \gtrsim 1$. This is just
the same region where the claimed Sommerfeld factor $S(v, \, \a) >>1$.

Can one escape the contradiction by appealing to small $\a_X$? It seems not: No small value for $\a_X$ is used in the logic of an ÒexactÓ
normalization citing Sommerfeld. Small coupling is also no protection from internal inconsistency. Unitarity and analyticity in perturbation
theory are exact facts maintained in a systematic way, order by order, regardless of the size of the coupling constant, small or large. When
violated, it shows the calculation was bad, just as indicated by sub-leading terms\footnote{Work by Cassel \cite{recents} and Iengo \cite{recents}
seek normalization pre-factors of the form $\sigma_{l} \ra \sigma_{l}S_{l}$ from the start. Those assumptions then contradict unitarity, as Cassel
noticed.}.

This problem of consistency is different from the one previously recognized. Dark matter interactions have finite range, while the infrared
singularities of re-summation come from infinite range. To account for this the authors of Ref.  \cite{arkani} argued that for a finite range
potential, a Sommerfeld enhancement would saturate when the deBroigle wavelength of colliding particles would be larger than the range of the
force.  A related statement actually follows from a $WKB$ approximation: when the de Broglie wavelength is tiny compared to the range, {\it and}
one works with wave functions at short distance, the range effects of a Yukawa potential $e^{- \mu r}/r $ drop out.  Note that the range criteria
do not depend on a coupling constant, and also don't specify any particular angular momentum channels.  Yet the singularities of scattering
amplitudes, and particularly Coulomb singularities, do depend on the couplings and angular momentum channels. Whatever the scale where analogies
between massless and massive models break down, the facts of partial wave unitarity are more general, and take precedence. They preclude large
enhancement in any particular channel of fixed angular momentum.

This leads to another useful bound. Replace every partial wave by the one with largest magnitude $|f_{max}|$. Sum them up: The result is \ba
\sigma \lesssim {4 \pi \over k^{2}} \sum^{\l_{max}} \, (2l+1)|f_{max}|^{2} = {4 \pi |f_{max}|^{2}(l_{max}+1)^{2} \over k^{2}}. \label{sigmax} \ea
This is a strict upper limit.  The Sommerfeld factor is not a resonance, and in the absence of resonances $|f_{max}|^{2}$ for every partial wave
is small in weakly coupled theories, making $\sigma$ small. The notion of canceling a perturbative factor of $\a_{X}^{2}$ with an enhancement of
$10^{4}$ (say) needs $l_{max} >> 10^{2}$ partial waves. This is difficult to conceive - or at least a high burden of proof - for finite-range,
perturbatively coupled Yukawa models. It is even more problematic that the angular momentum involved in annihilation is strictly limited by the
quantum numbers of intermediate states.  When the intermediate state consists of a single particle, the angular momentum is bounded by its spin,
closing the door on large spin sums.

Since each partial wave is finite, how do some Coulomb-dominated processes, such as Rutherford scattering, actually become singular as $v \ra 0$?
As Wigner \cite{wigner} and many others have noted, the Coulomb singularity is very special.  On semi-classical grounds (actually the facts of
Legendre series), in Eq. \ref{sigmax} the upper limit $l_{max} \sim r_{max}k$, where $r_{max}$ is the range of the potential. This gives \ba
\sigma \lesssim 4 \pi r_{max}^{2} |f_{max}|^{2} \lesssim 4 \pi r_{max}^{2} . \nn \ea The Coulomb singularity occurs because (1) the effective
range $r_{max} \ra \infty$, and (2) an infinite number of partial waves actually can contribute.  Closely related is Wigner's classic theorem
\cite{wigner} that power-law potentials $V(r) \gtrsim 1/r^{2}$ are needed to develop any kind of singularity.

\section{Concluding Remarks}

We have explored significant effects in halo annihilation rates due to natural widths of intermediate states. The problem is intricate due to
subtle interplay of energy scales. The Born approximation almost always fails, giving gross underestimates of reaction rates. Tiny values of
galactic halo velocities reverse an assumption that propagator widths might be ``small corrections.''  That perception comes from comparing widths
to particle masses, and does not capture the important features of halo annihilation.

Given that TeV-scale particles with typical electroweak type couplings may easily have $\Gamma/E_{0}>>1$, Breit-Wigner factors of ordinary
radiative corrections must generally be taken into account. Consistency of rates in particular channels, such as the apparent dominance of
leptons, still needs to be considered model by model. The fact that merely including basic physics of widths may be quite significant.  It revises
the basic picture of annihilation scenarios confronting the PAMELA-FERMI-PPB-BETS data in a positive way that increases the possibilities to find
new physics.

{\it Acknowledgments:} Research supported in part under DOE Grant Number DE-FG02-04ER14308. We thank Danny Marfatia, Doug McKay, Lou Cocke, Larry
Weaver  and Greg Adkins for useful comments. Courteous and diligent communications from Masahiro Ibe were helpful.

\end{document}